\documentclass[aps,pre,showpacs,
amssymb,twocolumn
]{revtex4}
\usepackage{graphicx}
\usepackage{amsmath}
\usepackage{amsfonts}
\usepackage{amssymb}
\usepackage{epsfig,libertine}
\usepackage[libertine]{newtxmath}
\usepackage{color}
\usepackage{dsfont, hyperref, xcolor}
\usepackage{comment}
\usepackage{enumerate}

\usepackage{amsthm} 
\newcommand{\bra}[1]{\langle#1\vert}
\newcommand{\ket}[1]{\vert#1\rangle}

\begin{document}

\title{Quantum advantage in batteries for Sachdev-Ye-Kitaev interactions}

\author{Gianluca~Francica}
\address{Dipartimento di Fisica e Astronomia, Universit\`{a} di Padova, via Marzolo 8, 35131 Padova, Italy}

\date{\today}

\begin{abstract}
A quantum advantage can be achieved in the unitary charging of quantum batteries if their cells are interacting. Here, we try to clarify with some analytical calculations whether and how this quantum advantage is achieved for sparse Sachdev-Ye-Kitaev (SYK) interactions {\color{black}and in general for fermionic interactions with disorder. To do this we perform a simple modelization of the interactions. In particular,} we find that for $q$-point rescaled sparse SYK interactions the quantum advantage goes as $\Gamma\sim N^{\frac{\alpha-q}{2}+1}$ for $q\geq\alpha\geq q/2$ and $\Gamma\sim N^{1-\frac{\alpha}{2}}$ for $q/2>\alpha\geq 0$,  where $\alpha$ is related to the connectivity and $N$ is the number of cells.
{\color{black} This shows how we can get $\Gamma\sim N$, i.e., an average power that scales as $N^2$ thanks to the disorder.} 
\end{abstract}

\maketitle

\section{Introduction}
Isolated quantum systems can be used to temporarily store some energy as quantum batteries, which can be later used {\color{black}to perform useful work} in a consumption center (see, e.g., Ref.~\cite{campaioli23} for a review). The extraction of the work can be performed by a cyclical change of the Hamiltonian parameters, and in this case the maximum average work extractable is called ergotropy~\cite{Allahverdyan04}.
In particular, by considering $N$ copies of the system, i.e., a quantum battery with $N$ cells, the total ergotropy can be larger than $N$ times the ergotropy of the single cell~\cite{Alicki13}. Interestingly, this amount of ergotropy can be also obtained without generating entanglement in the work extraction time-evolution~\cite{Hovhannisyan13}. In particular, work extraction from finite quantum systems can be related to some genuine quantum features, such as correlations~\cite{Francica17,Bernards19,Touil22,Francica22} and coherence~\cite{Francica20}. Furthermore, a more general optimization of the work extraction protocol can be performed by taking in account also the work fluctuations~\cite{Francica24,Francica242} with a quasiprobability distribution~\cite{Francicaqp22,Francicaqp222,Francicaqp23}.
The role of entanglement in charging quantum batteries was also investigated (see, e.g., Ref.~\cite{Gyhm24}) and, recently, a quantum charging distance was introduced in Ref.~\cite{Gyhm242}. 
Concerning the duration time of the protocol, the average power of a unitary charging of a battery with cells was originally investigated in Ref.~\cite{Binder15}, where the presence of interactions between the cells during the charging allows to achieve a so-called quantum advantage~\cite{Campaioli17,Gyhm22}. There are several proposals for the realization of quantum batteries, for instance by using
many-body interactions in spin systems~\cite{Julia-Farre20}, with a cavity assisted charging \cite{Ferraro18,Andolina19,Andolina18,Farina19,Zhang19,Andolina192,Crescente20}, in disordered chains~\cite{Rossini19} and fermionic Sachdev-Ye-Kitaev (SYK) interactions~\cite{Rossini20,Rosa20,Carrega21}, 
 to name a few.

{\color{black} A maximum quantum advantage can be achieved if the time-evolution during the charging process gives a geodetic, as originally shown in Ref.~\cite{Binder15}. However, it is not clear how this maximum quantum advantage can be achieved with many-body interactions. Here, with the aim to address this problem, we show how the maximum quantum advantage can be achieved if the charging process is performed with a SYK-like Hamiltonian, achieving a power $P$ that scales as $P\sim N^2$.} 
In particular, the SYK model~\cite{SachdevYe, kitaevtalk} has received a large interest in the last years (see, e.g., Ref.~\cite{Chowdhury22} for a review). Among its properties, it displays a resistivity that is linear with respect to the temperature, exhibiting a so-called Planckian transport~\cite{Chowdhury22,Patel19,Sachdev23}, a duality to a two-dimensional nearly anti-de Sitter space~\cite{kitaevtalk,kitaev18,maldacena16,sarosi19}, and many others, attracting the attention of both condensed matter and high energy physicists. 
Concerning its experimental realization, there are already some proposals, e.g., in solid state physics (see, e.g., Ref.~\cite{Franz18} for a review) and in cavity quantum electrodynamics platforms~\cite{hauke}.
Several studies investigate the mesoscopic physics by SYK model
(see, e.g., Refs.~\cite{song17,davison,Gnezdilov18,Can19,Guo20,Kulkarni22,Hosseinabadi23}).
Nonequilibrium dynamics was also investigated through this model, e.g., the heating of black holes at short times before the thermalization with a colder bath~\cite{Almheiri19,Zhang19,Cheipesh21}, the current driven by a double contact setup~\cite{Francica23}, and also a quantum quench~\cite{Eberlein17}. Other studies concern eternal traversable wormholes~\cite{Zhou20}, the Bekenstein-Hawking entropy~\cite{Kruchkov20} and  the existence of anomalous power laws in the temperature dependent conductance~\cite{Altland19}. Furthermore, the sparse SYK model was also investigated (see, e.g., Refs.~\cite{swingletalk,Garcia21,Caceres21,Caceres22,Tezuka23,Herasymenko23}).

Having discussed the importance of the SYK interactions, in the following we aim to clarify whether and how a quantum advantage in the power charging of quantum batteries can be achieved by exploiting these kinds of interactions. 
{\color{black}In order to do this we introduce some general preliminaries notions in Sec.~\ref{sec.preli}, showing how for extensive charging Hamiltonian the quantum advantage can scale at most linearly with $N$. Then, in Sec.~\ref{sec.model}, we explain how this linear scaling can be achieved thanks to quenched disorder when the interactions are SYK interactions.
Basically, since the SYK maximum energy goes as the square root of the number of fermions, we have to rescale the Hamiltonian to obtain full extensibility if we want to achieve maximum quantum advantage.
Then, we proceed by introducing a simple model in Sec.~\ref{sec.analexp}, in order to investigate further the problem. 
}Finally, we discuss further our results in Sec.~\ref{sec.conclusion}.

\section{Quantum advantage}\label{sec.preli}
Let us discuss the quantum advantage in general. 
In order to derive {\color{black}a bound for} the scaling of the quantum advantage, we focus on a quantum battery having $N$ cells (or copies) and Hamiltonian $H_0 = \sum_{i=1}^N h_i$, where $h_i$ is the Hamiltonian of the single cell. The battery is initially prepared in the ground-state $\ket{\psi(0)}=\ket{0}^{\otimes N}$ having zero energy, i.e., $h_i\ket{0}=0$ for all $i=1,\ldots,N$. The charging process leads the system to the final state $\ket{\psi(\tau)}=U_{\tau,0}\ket{\psi(0)}$ where $U_{t,0}$ is the time-evolution unitary operator generated by the time-dependent Hamiltonian $H(t)$ with $t\in[0,\tau]$ such that $H(0)=H(\tau)=H_0$. The time-evolved state $\ket{\psi(t)}$ with $t\in[0,\tau]$ defines a curve $\mathcal C$ in the projective space having Fubini-Study length~\cite{Anandan90}
\begin{equation}\label{eq.FB}
l(\mathcal C)= \int_0^\tau \Delta H(t) dt\,,
\end{equation}
where $\Delta H^2(t) = \langle H^2(t)\rangle_t-\langle H(t)\rangle_t^2$ and the averages are calculated with respect to the state $\ket{\psi(t)}$, i.e., we defined $\langle X \rangle_t = \bra{\psi(t)} X \ket{\psi(t)}$ for any operator $X$.
In this {\color{black}paper, for simplicity} we focus on a charging process generated by performing two quantum quenches at the initial and final times $t=0,\tau$.
{\color{black}In detail}, the time-dependent Hamiltonian is defined such that $H(t)=H_0$ for $t=0,\tau$ and $H(t)=H_1$ for $t\in (0,\tau)$, with $[H_1,H_0]\neq 0$ in order to achieve a  non-trivial dynamics. In this case, the unitary time-evolution operator is $U_{t,0}=e^{-i H_1 t}$ when $t\in(0,\tau)$ and the Fubini-Study length in Eq.~\eqref{eq.FB} simplifies to
\begin{equation}\label{eq.FBquench}
l(\mathcal C) = \Delta H_1 \tau\,,
\end{equation}
where
\begin{equation}\label{eq.deltaH1def}
\Delta H_1^2 = \langle H_1^2\rangle_0-\langle H_1\rangle_0^2\,.
\end{equation}
We note that $\Delta H_1^2$ is equal to the work variance of the first quench at $t=0$, i.e., it can be obtained as $\Delta H_1^2 = - G''(0)$, where $G(u)=\ln \langle e^{i H_1 u}\rangle_0$ is the cumulant generating function of the work done in the first quench.

{\color{black}Thus, in general the duration time $\tau$ of the unitary charging process is obtained from the Fubini-Study length $l(\mathcal C)$ and the energy uncertainty $\Delta H_1$ through Eq.~\eqref{eq.FBquench}. This shows how to optimize the duration time $\tau$ of the charging process, we can choose a Hamiltonian $H_1$ that minimizes the length $l(\mathcal C)$ and maximizes $\Delta H_1$ with the constraint that the work scales linearly with $N$. It is known~\cite{Binder15} that we can get a minimum length $l(\mathcal C)\sim\mathcal O(1)$ and $\Delta H_1 \sim N$ with a work that scales linearly with $N$ when the curve $\mathcal C$ is a geodetic, achieving the scaling $\tau \sim 1/N$ (as also explained later in the paper).}
To define a quantum advantage, {\color{black}as in Ref.~\cite{Binder15}} we always choose $H_1$ such that the gap $\Delta E_1 = E_{max}-E_{min}$ between its maximum and minimum eigenvalues, which are $E_{max}$ and $E_{min}$, is not larger than the gap $\Delta E_0$ between the maximum and the minimum eigenvalues of $H_0$.
Then, by considering $H_0$ with a maximum eigenvalue that linearly scales with $N$, we get  $\Delta E_1 \leq \Delta E_0 \sim N$.
{\color{black}The variance $\Delta H_1^2$ is bounded by} the Bhatia-Davis inequality {\color{black}(originally introduced in Ref.~\cite{Bathia00})}
\begin{equation}\label{eq.BhatiaDavis}
\Delta H_1^2\leq (E_{max}-\mu)(\mu-E_{min})\,,
\end{equation}
where $\mu=\langle H_1\rangle_0$, so that an optimal charging process can be achieved when the Bhatia-Davis inequality is saturated.
{\color{black}To prove Eq.~\eqref{eq.BhatiaDavis}, it is enough to note that $H_1\geq E_{min}$ and $E_{max}\geq H_1$, where $X\geq Y$ means that the operator $X-Y$ is semidefinite positive. Then, by noting that for any $X\geq 0$ and $Y\geq 0$ such that $[X,Y]=0$, we get $XY\geq 0$, we have
\begin{eqnarray}
0 &\leq& \langle (E_{max}-H_1)(H_1 - E_{min})\rangle_0 \\
&=& \mu E_{max} + \mu E_{min}  - E_{max}E_{min}-\langle H_1^2\rangle_0\,,
\end{eqnarray}
from which we easily get Eq.~\eqref{eq.BhatiaDavis} by noting that $\Delta H_1^2$ is given by Eq.~\eqref{eq.deltaH1def}. }
{\color{black} From Eq.~\eqref{eq.BhatiaDavis},} we deduce that when
\begin{equation}\label{eq.delta1scala}
\Delta H_1^2\sim N^{2a}\,,
\end{equation}
we get $a\leq 1$. To show it, it is enough to
consider $E_{min}=0$, so that we can get at most $E_{max}\sim N$ and $\mu\sim N$. 
{\color{black}In particular, the average value $\mu$ scales at most as $\mu \sim N$, i.e., the Hamiltonian $H_1$ is at most extensive, since $\mu$ is achieved as a convex combination of the energy levels, which lie in the interval $[E_{min},E_{max}]$. {\color{black} Of course, in general we get the bound $a\leq b$, if the right-hand side of the Bhatia-Davis inequality scales as $(E_{max}-\mu)(\mu-E_{min})\sim N^{2b}$ with $b\leq 1$.} Clearly, in order to maximize $\Delta H_1$ we  consider the optimal situation of a full extensive Hamiltonian $H_1$ having $\Delta E_1 \sim N$.}

In order to define a quantum advantage, let us examine two different situations. 
In the first one {\color{black}we perform a parallel charging where} the duration time is denoted with $\tau^{\|}$ and the charging process is obtained with $H_1 =  \sum_{i=1}^N v_i $ such that $\Delta E_1\sim N$, where $v_i$ is a local operator of the i-th cell, so that there are no interactions among the cells. {\color{black} In this paper, we refer to any non-local term, e.g., the term $v_i \otimes v_j$ with $i\neq j$, as interaction (among the cells).}
The average power is $P^{\|}=W^{\|}/\tau^{\|}$, where $W^{\|} = \langle H_0 \rangle_{\tau^{\|}}$ is the average work, which scales as $W^{\|}\sim N$ when $\tau^{\|}\sim O(1)$.
{\color{black} To show it, it is enough to note that in this case $W^{\|} = \sum_i \bra{0}e^{i v_i \tau^{\|}}h_ie^{-i v_i \tau^{\|}}\ket{0} $, which of course scales as $N$ if $\tau^{\|} \sim O(1)$ and $[v_i,h_i]\neq 0$. }
In contrast, in the second situation the duration time is denoted with $\tau^{\sharp}$ and interactions are allowed during the charging process, e.g., terms like $\sum_{i_1,i_2, \ldots, i_q} v_{i_1} v_{i_2}\cdots v_{i_q}$ may be present in the Hamiltonian $H_1$, {\color{black}where we dropped the $\otimes$'s for brevity}.
{\color{black}In this second situation the only constraint to choose $H_1$ is that the gap $\Delta E_1$ scales at most linearly with $N$. The duration time $\tau^\sharp$ is such that the work scales as $W^\sharp \sim N$. In order to discuss the scaling for the average work $W^\sharp$, which reads
$W^\sharp = \sum_i \langle e^{iH_1\tau^\sharp}h_ie^{-iH_1\tau^\sharp}\rangle_0$, we use the Baker-Campbell-Hausdorff formula and we get the power series
\begin{equation}
W^\sharp=\sum_i \langle H_1h_iH_1\rangle_0{\tau^\sharp}^2+\cdots = \sum_k\sum_i w_{ik} {\tau^\sharp}^k \,,
\end{equation}
where $w_{i2}= \langle H_1h_iH_1\rangle_0$. Let us focus on a duration time $\tau^\sharp=\tau'/N^z$, with $\tau'\sim \mathcal{O}(1)$. In this case, if $w_{ik}\sim N^{kz}$ for all $i$ and some $k$, then we get $W^\sharp\sim N$ for some $\tau'$. Thus, the work $W^\sharp$ scales linearly with $N$ for a duration time $\tau^\sharp\sim 1/N^z$. Of course, the exponent $z$ depends on the chosen Hamiltonian $H_1$ and can be at most equal to $a$, $z\leq a$.}
{\color{black} Depending on the interaction} we can get {\color{black}the optimal situation with} an average work $W^{\sharp}\sim N$ for a length $l(C)\sim O(1)$, i.e., from Eq.~\eqref{eq.FBquench}, for a duration time $\tau^{\sharp} \sim 1/\Delta H_1$. 
{\color{black}From the above discussion, this scaling is achieved if $\langle H_1h_iH_1\rangle_0\sim \Delta H_1^2$ for all $i$, which implies that $z=a$ if $\Delta H_1^2$ scales as in Eq.~\eqref{eq.delta1scala}. }
{\color{black} To explicitly show how this situation can be achieved with some $H_1$, we follow Ref.~\cite{Binder15}. Thus, we consider the charging process achieved with the Hamiltonian $H_1 = N \lambda (\ket{E^0_{max}}\bra{E^0_{min}}+\ket{E^0_{min}}\bra{E^0_{max}})$, where $\ket{E^0_{max}}$ is the eigenstate of $H_0$ with maximum energy and $\ket{E^0_{min}}=\ket{0}^{\otimes N}$.
Since, the initial state is $\ket{\psi(0)}=\ket{E^0_{min}}$, the time-evolved state is $\ket{\psi(\tau^\sharp)} = \cos(l(C)) \ket{E^0_{min}} + \sin(l(C)) \ket{E^0_{max}}$, and the curve  $\mathcal C$ is a geodetic having a length  $l(C)= N\lambda \tau^\sharp$. Thus, when $W^{\sharp}\sim N$, we get  $l(\mathcal C) \sim \mathcal O(1)$, and thus $\tau^\sharp \sim 1/N$. In particular, we also get $\tau^\sharp \sim 1/N$ from Eq.~\eqref{eq.FBquench}
by noting that in this case $\Delta H_1 \sim N$. It is easy to check that the sufficient condition, $\langle H_1h_iH_1\rangle_0\sim \Delta H_1^2$ for all $i$, to get $\tau^{\sharp} \sim 1/\Delta H_1$ is satisfied for this particular Hamiltonian $H_1$. However, in general we can also get some interactions among the cells described by $H_1$ with an average value $\mu$ that is sub-extensive, e.g., such that $\lim_{N\to \infty} \mu/N =0$, although $\Delta E_1 \sim N$. Thus, in general, we can consider $\Delta H_1 \sim N^a$ with $a$ bounded by the  Bhatia-Davis inequality in Eq.~\eqref{eq.BhatiaDavis}, such that $a\leq 1$.} 
In the optimal case the time-evolution connects the initial and final states with a geodetic having a length $l(C)=\pi/2$, so that the final state has maximal energy, i.e., $\ket{\psi(\tau^\sharp)}=\ket{E^0_{max}}$.
{\color{black}In the second situation 
the average power is $P^{\sharp}=W^{\sharp}/\tau^{\sharp}$, and if $W^{\sharp}=W^{\|}$, 
we get the quantum advantage~\cite{Campaioli17}}
\begin{equation}
\Gamma=\frac{P^{\sharp}}{P^{\|}} = \frac{\tau^{\|}}{\tau^{\sharp}} \,,
\end{equation}
{\color{black}so that, since the optimal Fubini-Study length scales as $l(\mathcal C)\sim \mathcal O(1)$, for a given $\Delta H_1$ as in Eq.~\eqref{eq.delta1scala}, we get the general bound
\begin{equation}
\Gamma \lesssim N^a\,,
\end{equation}
from which, when $l(\mathcal C) \sim \mathcal O(1)$ the quantum advantage} 
 scales as
\begin{equation}\label{eq.gammaa}
\Gamma\sim N^{a}\,.
\end{equation}
In particular, since $a\leq 1$, we get the {\color{black}general} bound
\begin{equation}\label{eq.bound}
\Gamma \lesssim N\,,
\end{equation}
{\color{black}which is difficult to saturate. 
Let us show how it can be saturated in a system of fermions with quenched disorder.}

\section{Role of disorder}\label{sec.model}
{\color{black} We focus on a local Hilbert space (of a single cell) having dimension $d=2$. Thus, by following Ref.~\cite{Rossini20},
we consider $h_i=\epsilon_0(\sigma^y_i+1)$, where $\sigma^x_i$, $\sigma^y_i$ and $\sigma^z_i$ are the local Pauli matrices of the i-th cell. Of course, the 
parallel charging process, which gives the power $P^{\|}$, can be performed for instance with the local operator $v_i=\lambda_0\sigma^x_i$, or any other operator $v_i$ that does not commutate with $h_i$.}
{\color{black} Now, let us focus on a charging process with many-body interactions that gives $P^\sharp$.
To briefly illustrate the main concept, we start to consider a Hamiltonian $H_1$ which can be expressed as a disorder-free quadratic Hamiltonian in some fermionic operators. In order to have $\Delta E_1\sim N$ extensive, we consider
\begin{equation}\label{eq.Hclean}
H_1 = i J_0 \sum_{i>j} \gamma_i \gamma_j\,,
\end{equation}
where $\{\gamma_i,\gamma_j\}=2\delta_{i,j}$ and $J_0\sim 1/\sqrt{N}$. In particular, the energy scale $J_0$ guarantees that $\Delta E_1\sim N$ and thus the model is extensive. In this case $\Delta H_1 \sim \sqrt{N}$, thus we get $a=1/2$ and there is no way to saturate the bound for the quantum advantage of Eq.~\eqref{eq.bound}, which can be achieved only if $a=1$.
In detail, the operators $\gamma_i$ are related to the Pauli matrices $\sigma^x_i$, $\sigma^y_i$ and $\sigma^z_i$ through some transformation, e.g., through the Jordan-Wigner transformation
\begin{eqnarray}
\label{eq.JW1}\gamma_{2l}&=&\left(\prod_{k=1}^{l-1} \sigma^z_k\right) \sigma^x_l\,,\\
\label{eq.JW2}\gamma_{2l-1}&=&\left(\prod_{k=1}^{l-1} \sigma^z_k\right) \sigma^y_l\,.
\end{eqnarray}
In contrast, if we consider quenched disordered couplings among the fermions, from Eq.~\eqref{eq.Hclean} we get the Hamiltonian
\begin{equation}\label{eq.H1disoq2}
H_1 = i \sum_{i>j} J_{ij} \gamma_i \gamma_j\,,
\end{equation}
where $J_{ij}\sim \mathcal O(1)$ random with zero average in order to have the optimal scaling $\Delta E_1 \sim N$. Furthermore, for this Hamiltonian $H_1$ we easily get that the disorder average of $\Delta H_1^2 $ scales as $N^2$, and thus $a=1$.
{\color{black} Since $\sigma^y_i=\left(\prod_{l=1}^{i-1} i\gamma_{2l-1}\gamma_{2l}\right)\gamma_{2i-1}$ is a product of an odd number of $\gamma$ operators and the terms in the Hamiltonian $H_1$ are products of an even number of $\gamma$ operators, we get $\langle H_1h_iH_1\rangle_0\sim \Delta H_1^2$ for all $i$ by noting that $h_i=\epsilon_0(\sigma^y_i+1)$ and $\sigma^y_iH_1= - H_1 \sigma^y_i$.}
This 
 implies that we can saturate the bound for the quantum advantage of Eq.~\eqref{eq.bound} by using quenched disorder.
Let us investigate this idea in the case of $q$-point  sparsed interactions.

In order to introduce the Hamiltonian $H_1$,}
we start to consider 
$q$-point sparse fermionic interactions, i.e., $H_1 = H_{SYK,q}$ with
\begin{equation}\label{eq.SYKq}
H_{SYK,q} = i^{\frac{q}{2}} \sum_{1\leq i_1 < i_2 < \ldots < i_q\leq 2N} x_{i_1 i_2 \ldots i_q}j_{i_1 i_2 \ldots i_q} \gamma_{i_1}\gamma_{i_2}\cdots \gamma_{i_q}\,,
\end{equation}
where {\color{black} $q$ is an even positive integer,} 
 and the quenched disorder $j_{i_1 i_2 \ldots i_q}$ has a Gaussian distribution with zero mean and variance $j^2 (q-1)!/N^{q-1}$, where $j\sim \mathcal O(1)$. We consider $x_{i_1 i_2 \ldots i_q}=1$ with probability $p$ and $x_{i_1 i_2 \ldots i_q}=0$ with probability $1-p$, defining the connectivity of the fermions.
We {\color{black} can calculate} the average number of $q$-point connections {\color{black} $C_q$ as the average number of nonzero couplings so that}
\begin{equation}
C_q =\sum_{1\leq i_1 < i_2 < \ldots < i_q\leq 2N} \overline{x_{i_1 i_2 \ldots i_q}} \sim p N^q\,,
\end{equation}
where the bar denotes the average over the disorder.
For our purposes, we focus on
\begin{equation}\label{eq.alpha}
C_q \sim N^\alpha\,,
\end{equation}
with $0\leq \alpha\leq q$.
We note that the maximum eigenvalue of the Hamiltonian in Eq.~\eqref{eq.SYKq} scales as $E_{max}\sim \sqrt{N}$ {\color{black}for $\alpha=q$}, so that the bound of Eq.~\eqref{eq.bound} cannot be saturated, since from Eq.~\eqref{eq.BhatiaDavis} we get {\color{black}at most $\Delta H^2_1 \sim N$, and thus $a=1/2$. 
In particular, {\color{black} we recall that} numerical investigations done for $q=4$ in Ref.~\cite{Rossini20} indicate that the power scales as $P^\sharp\sim N^{\frac{3}{2}}$ and of course $a=1/2$ if $W^\sharp\sim N$.}
{\color{black} By performing a rescaling of both the spectrum $H_1 \mapsto M H_1$ and  the duration time $\tau^\sharp \mapsto \tau^\sharp/M$, the length $l(\mathcal C)$ remains unchanged, where $M$ is a positive constant. }
Then, in order to get $a>1/2$ we rescale the energy spectrum, i.e., we define
{\color{black}
\begin{equation}\label{eq.SYKre}
H_1= M  H_{SYK,q}\,.
\end{equation}
For $\alpha=q$, $M=\sqrt{N}$ in order to get the optimal scaling $E_{max}\sim N$, so that we can get $a=1$ and so we can saturate} the bound in Eq.~\eqref{eq.bound}, since the fermions can be fully-connected. 
{\color{black}For instance, for $q=2$ we get a Hamiltonian as that in Eq.~\eqref{eq.H1disoq2}. We note that this Hamiltonian can be obtained from $H_1=\sum_i \lambda_i (i \chi_{2i}\chi_{2i-1}-1)$ by considering $\chi_i=\sum_j W_{ij} \gamma_j$, where $W_{ij}$ is some $2N\times 2N$ random orthogonal matrix.}

In {\color{black}the next section}, we aim to clarify the role of the interactions and connectivity for the saturation of the bound through a suitable simplification of the Hamiltonian $H_1${\color{black}, by determining the optimal quantum advantage achievable for given $\alpha$ and $q$.}
{\color{black}To do this, we firstly
calculate how the gap $\Delta E_1$ scales with $N$ for $q=2$ if $M=\sqrt{N}$, so that from Eq.~\eqref{eq.SYKre} we get $H_1 = i \sum A_{ij}\gamma_i \gamma_j$, {\color{black} where $A_{ij}\sim \mathcal O(1)$}. By randomly generating $x_{ij}$, the nonzero entries of the skew-symmetric matrix $A_{ij}$  are almost uniformly distributed, thus the disorder average of $\Delta E_1$ is determined by only these matrices. Let us consider $C_2\sim N^\alpha$ as in Eq.~\eqref{eq.alpha}. For $1\leq \alpha \leq 2$, all the $2N$ fermions appear in the Hamiltonian $H_1$, so that $\Delta E_1 \sim N$. In contrast, for $0\leq \alpha\leq 1$, only $\mathcal O(N^\alpha)$ fermions appear in the Hamiltonian $H_1$, and thus $\Delta E_1 \sim N^\alpha$.
{\color{black}Then, for $q=2$, in order to get a gap $\Delta E_1 \sim N$, we can consider the multiplying factor  $M=\sqrt{N}$ for $\alpha \geq 1$ and $M=N^{\frac{1}{2}+1-\alpha}$ otherwise. }

\section{Simplified model}\label{sec.analexp}
{\color{black}Based on the previous discussion}, we introduce {\color{black}an extensive} simplified model for $q=2$ as $H_1 = V$ with
\begin{equation}\label{eq.Vclean}
V= N^{x\theta(x)}\left(i\sum_{i,j}y_{ij}\lambda_{ij} \gamma_i \gamma_j + w N\right)
\end{equation}
and from $V$ we get $q$-point interactions by considering $V^k$ with $k=q/2$, so that the simplified Hamiltonian for arbitrary $q$ reads
\begin{equation}\label{eq.H1}
H_1= (J V)^k\,,
\end{equation}
where $\theta(x)$ is the Heaviside step function defined as $\theta(x)=1$ if $x\geq 0$ and $\theta(x)=0$ otherwise, $y_{ij}=1$ with probability $p_1$ and $y_{ij}=0$ with probability $1-p_1$, where $p_1$ is such that $p_1^k\sim p$, so that $\sum \overline{y_{i_1i_2}\cdots y_{i_{q-1}i_q}}\sim C_q$ and $x\in[-1,1]$ is such that $p_1\sim 1/N^{x+1}$, i.e., $x=1-2\alpha/q$.
Furthermore, we introduced the couplings $\lambda_{ij}\sim O(1)$ such that $\overline{\lambda_{ij}}=0$, a c-number $w$ that scales as $p_1$, i.e., $w\sim p_1$, and 
 an energy scale $J$ such that we get the optimal scaling $\Delta E_1 \sim N$ for any positive integer $k$. 
Since for $k=1$, the gap scales as $\Delta E_1 \sim N$, we consider the coupling $J= N^{-\frac{k-1}{k}}$.
{\color{black} In order to explain in detail why $\Delta E_1 \sim N$ for any positive integer $k$, we start to note that the disorder average of the gap of the Hamiltonian $H_1=V$ scales linearly with $N$ for any $\alpha\in[0,q]$ due to the multiplying factor $N^{x\theta(x)}$ added in Eq.~\eqref{eq.Vclean}.
Furthermore, any exponentiation of $V$ is exactly diagonalizable, since $V$ is quadratic and can be exactly diagonalized by performing a linear 
transformation (see, e.g., Ref.~\cite{Lieb61}), and thus the simplified model in Eq.~\eqref{eq.H1} can be exactly diagonalized in the same way.
Then, since for a random realization of the disorder, the gap of $H_1=V$ scales linearly with $N$, the disorder average of the maximum eigenvalue of $V^k$ scales as $N^k$, and we must multiply $V^k$ by something that scales as $N^{1-k}$ in order to get a gap that scales linearly with $N$.} 
%

{\color{black}We can now proceed with the calculation of the scaling for the quantum advantage for this model.
Since the gap is extensive, the scaling for the quantum advantage for given $\alpha$ and $q$ is the maximum one.
A key quantity is the disorder average of the uncertainty $\Delta H_1^2$.
One can calculate the average over the disorder of $\Delta H_1^2$ by performing the disorder average of Eq.~\eqref{eq.deltaH1def}, as done below in Eq.~\eqref{eq.SYKdisoave}. However, the same result can be also achieved from the disorder average of the cumulant generating function $G(u)$.
For this reason, we will try to calculate the generating function $G(u)$ for this type of models in the following section.
Summarizing, having defined $H_1$, before to achieve the optimal quantum advantage we perform a calculation separated in two independent chunks: we firstly perform a calculation of the generating function in Sec.~\ref{sec.cumu}, then we calculate the scaling of the disorder average of $\Delta H_1^2$ in Sec.~\ref{sec.deltaH1}. The quantum advantage is calculated from $\Delta H_1^2$ and discussed in Sec.~\ref{sec.qadva}.}

\subsection{Cumulant generating function $G(u)$}\label{sec.cumu}
Let us
derive a general result, which holds for a generic operator $V$, not necessarily of the form in Eq.~\eqref{eq.Vclean}.
In this case we get
\begin{equation}\label{eq.momgen}
\langle e^{iu H_1}\rangle_0 = \langle e^{iu(JV)^k} \rangle_0\,.
\end{equation}
Since we are interested only to the variance $\Delta H_1^2$, we aim to calculate Eq.~\eqref{eq.momgen} for $u\to 0$. We search a function $f(z)$ such that
\begin{equation}\label{eq.genHS}
 e^{iux^k}  \sim \int_{C} dz e^{i f(z)/u + i z x }
\end{equation}
as $u\to 0$, where $C$ is a stationary phase path so that the dominant contribution will be given by the stationary point of $f(z)/u + z x$. We get
\begin{equation}
\frac{f(z)}{u} = c \left( \frac{z^k}{u}\right)^{\frac{1}{k-1}}\,,
\end{equation}
where $c$ is a number that does not depend on $u$. For instance, for $k=2$ Eq.~\eqref{eq.genHS} reduces to the Hubbard-Stratonovich transformation. Thus, we get
\begin{equation}\label{eq.1}
\langle e^{iu H_1}\rangle_0 \sim \int_{C} dz e^{i f(z)/u}\langle e^{izJV} \rangle_0\,.
\end{equation}
As $N\to \infty$, to do some calculation we consider the generating function
\begin{equation}\label{eq.gf1}
\langle e^{izJV} \rangle_0 \sim e^{i N^{\frac{1}{k}} g(z)}\,,
\end{equation}
where $g(z)$ is a certain function.
We consider that the stationary point $z$ approaches to zero as $u\to 0$, thus we can consider the Taylor expansion $g(z) = g'(0) z + g''(0) z^2/2 + \cdots$ in the integral in Eq.~\eqref{eq.1}. By calculating the stationary point up to $\mathcal O(u^3)$ terms, we get
\begin{equation}\label{eq.asy}
\langle e^{iu H_1}\rangle_0 \sim \exp\left(i u {g'}^k(0) N + c'_1 u^2 {g'}^{2k-2}(0)g''(0) N^{\frac{2k-1}{k}} + \cdots\right)\,,
\end{equation}
where $c'_l$ are numbers that do not depend on $u$ and different from $c$.
{\color{black} The cumulant generating function $G(u)$ is the argument of the exponential function in Eq.~\eqref{eq.asy} by definition, 
 and in principle the terms in the ellipsis can be calculated by performing an asymptotic expansion.} 

\subsection{Scaling of $\Delta H_1^2$ with $N$}\label{sec.deltaH1}
{\color{black}From} Eq.~\eqref{eq.asy} we get the general result
\begin{equation}\label{eq.deltaH1square}
\Delta H_1^2 \sim \sum_{l=1}^{k} N^{\frac{2k-l}{k}} c'_l {g'}^{2k-2l}(0){g''}^l(0)  \,.
\end{equation}
{\color{black}The result in Eq.~\eqref{eq.deltaH1square} can be also easily derived by noting that the variance $\Delta H_1^2$ in Eq.~\eqref{eq.deltaH1def} is explicitly expressed in terms of the 2k-th and k-th moments of the operator $J V$, which are $\langle X^{2k}\rangle_0 $ and $\langle X^k \rangle_0$, where $X=JV$. By neglecting all the derivatives of $g(u)$ higher than the second one, i.e., by considering $g(z) = g'(0) z + g''(0) z^2/2 + \cdots$, from Eq.~\eqref{eq.gf1}, the operator $X$ is like a Gaussian variable, which is always true for small-deviations if the variance of $X$ is non-zero. Then, the moments $\langle X^{2k}\rangle_0 $ and $\langle X^k \rangle_0$ are related to the first two cumulants of $X$, which are $\mu = N^{\frac{1}{k}}g'(0)$ and $\sigma^2 =  -i N^{\frac{1}{k}} g''(0)$, by the general expansion of Eq.~\eqref{eq.deltaH1square}. This confirms the calculation done in the previous section.}

{\color{black}From an explicit calculation, } for $V$ of Eq.~\eqref{eq.Vclean} we have
\begin{equation}
g'(0)\sim N^{x\theta(x)}\left(\frac{1}{N}\sum_{i\neq j} y_{ij}\lambda_{ij}C_{ij}+w \right)
\end{equation}
and
\begin{equation}
ig''(0)\sim N^{2x\theta(x)+1} \frac{J}{N^2}\sum_{i\neq j}\sum_{k\neq l} y_{ij}y_{kl}\lambda_{ij}\lambda_{kl} C_{il}C_{jk}\,,
\end{equation}
{\color{black} where we used the Wick theorem to obtain the above expression of $g''(0)$ }
with $C_{ij}=i\langle \gamma_i \gamma_j\rangle_0$. 
Then, for $\overline{\lambda_{ij}}=0$ we can consider $g'(0)\sim N^{x\theta(x)} w$, and we get
\begin{equation}\label{eq.Kdef}
 \Delta H_1^2 \sim J N^{2k x \theta(x)+1} \sum_{l=1}^{k}c_l \Lambda_{2l} w^{2k-2l} = \sum_{l=1}^{k} K_{2l}\,,
\end{equation}
where we defined
\begin{eqnarray}\label{eq.Lambdan}
\nonumber \Lambda_n&=&\frac{1}{N^{n}}\sum_{i_1\neq j_1} \cdots \sum_{i_{n}\neq j_{n}} y_{i_1j_1}\cdots y_{i_{n} j_{n}}\lambda_{i_1j_1}\cdots \lambda_{i_{n} j_{n}}\\
&& \times C_{i_1j_2} C_{i_2j_1} \cdots C_{i_{n-1}j_{n}}C_{i_{n} j_{n-1}}\,.
\end{eqnarray}
The scaling of $\Delta H_1^2$ with $N$ is obtained by averaging over the disorder, and we can get different scalings depending on the probability distributions of $\lambda_{ij}$ and $w$.
If the moment $\overline{\lambda^{2}_{ij}}$ is nonzero we get $\overline{\Lambda}_{2l}\sim p_1^{l}$ for $l=1,\ldots,k$, and $\overline{K_{2l}}$ scales as
\begin{equation}\label{eq.Cnn}
\overline{K_{2l}} \sim N^{qx\theta(x)+\frac{2}{q}+\frac{2(\alpha-q)(q-l)}{q}}\,.
\end{equation}
Then, we get 
\begin{equation}\label{eq.sumdeltah1}
\Delta H_1^2 \sim N^{qx\theta(x)+2}\sum_{l=1}^{\frac{q}{2}}c_{l} N^{\frac{2(\alpha-q)(q-l)}{q}}\,,
\end{equation}
where $c_l\neq 0$ if $\overline{w^{2k-2l}}\neq 0$. We note that for correlated disorder $c_l=0$ if $\overline{\lambda_{i_1j_1}\cdots \lambda_{i_{2l} j_{2l}}}= 0$, e.g., if $\lambda_{ij}=z\tilde\lambda_{ij}$ with $\overline{z^{2l}}=0$.
\subsubsection{Rescaled SYK interactions}
For the SYK interactions we can calculate the scaling of $\Delta H_1^2 $ directly from the average over the disorder of Eq.~\eqref{eq.deltaH1def}, e.g., from Eq.~\eqref{eq.SYKq}, the disorder average of $H_{SYK,q}^2$ reads
\begin{eqnarray}\label{eq.SYKdisoave}
\nonumber \overline{H^2_{SYK,q}} &=& (-1)^{\frac{q}{2}} \frac{pj^2 (q-1)!}{N^{q-1}}\sum_{1\leq i_1 < \ldots < i_q\leq 2N}\gamma_{i_1}\cdots \gamma_{i_q}\gamma_{i_1}\cdots \gamma_{i_q}\\
 &\propto& p j^2 N
\end{eqnarray}
and the disorder average of $\langle H_{SYK,q}\rangle_0^2$ scales as $pN^{1-\frac{q}{2}}$, from which from Eq.~\eqref{eq.SYKre} with $M=N^{\frac{qx\theta(x)}{2}+\frac{1}{2}}$, we easily obtain 
\begin{equation}
\Delta H_1^2 \sim N^{qx\theta(x)+2+\alpha-q}\,,
\end{equation}
which is the scaling corresponding to the term with $l=q/2$ in the sum of Eq.~\eqref{eq.sumdeltah1}. 
{\color{black} We note that the scaling achieved for $\Delta H_1^2$ is optimal, since it tends to saturate the Bhatia-Davis inequality in Eq.~\eqref{eq.BhatiaDavis}. }
{\color{black} In particular, we can easily understand the factor $M=N^{\frac{qx\theta(x)}{2}+\frac{1}{2}}$ by noting that the SYK Hamiltonian in Eq.~\eqref{eq.SYKq} multiplied by this value of $M$ has the same form of the $q$-point interactions term in the simplified model of Eq.~\eqref{eq.H1}.}

\subsection{Discussion of the quantum advantage}\label{sec.qadva}

Thus, 
if for ${\tau^\sharp}^2 \sim 1/\Delta H_1^2$ the average work scales as $W^\sharp \sim N$, {\color{black} then of course from Eq.~\eqref{eq.FBquench} the Fubini-Study length scales as $l(C)\sim \mathcal O(1)$ and} from Eq.~\eqref{eq.gammaa} the quantum advantage can scale in different ways, which are
\begin{equation}\label{eq.gamman}
\Gamma \sim N^{\frac{qx\theta(x)}{2}+1+\frac{(\alpha-q)(q-l)}{q}}
\end{equation}
with $l=1,\ldots,q/2$. 
{\color{black}This means that the power scales as $P^\sharp \sim N \Gamma$.}
For instance, we note that $W^\sharp = \langle H_1 H_0 H_1 \rangle_0 {\tau^\sharp}^2 + \cdots$ and from a disorder average we get $\langle H_1 h_i H_1 \rangle_0\sim \Delta H_1^2$ for certain operators $h_i$ {\color{black}(e.g., for $h_i = \epsilon_i (\sigma^y_i+1)$ and $\gamma_i$ given by the Jordan-Wigner transformation in Eqs.~\eqref{eq.JW1}-\eqref{eq.JW2})}, thus by summing we get $W^\sharp \sim N$. 

Then, from the simplified model we get different classes of Hamiltonian operators with at most $q$-point interactions with disorder exhibiting the scalings of Eq.~\eqref{eq.gamman} and the rescaled SYK belongs to one of these classes. 
In particular, the Hamiltonian operators of Eq.~\eqref{eq.H1} can be considered as the representative elements of the classes. From a physical point of view, we note that the classes are a consequence of a connectivity given by the product $x_{i_1\ldots i_q} = y_{i_1i_2}\cdots y_{i_{q-1}i_{q}}$ and interactions having an order lower than $q$ coming from the exponentiation of $V$.
For the SYK class, for $x\in[-1,0]$ we get
\begin{equation}\label{eq.final}
\Gamma\sim N^{\frac{\alpha-q}{2}+1}
\end{equation}
for $\alpha\geq q/2$. In particular, we note that, while for $p\sim 1/N^2$ we get $q-\alpha= 2$ and there is no quantum advantage, the bound in Eq.~\eqref{eq.bound} is saturated for $\alpha=q$.
On the other hand, for $x\in(0,1]$ we get
\begin{equation}\label{eq.final2}
\Gamma\sim N^{1-\frac{\alpha}{2}}
\end{equation}
for $q/2>\alpha\geq 0$. Then, the bound in Eq.~\eqref{eq.bound} is saturated for $\alpha=0$, e.g., {\color{black}with only one $q$-point connection {\color{black} in average}}.
We note that the case $\alpha=0$ can be easily mapped into the case $\alpha=q$. To do this, we express the Hamiltonian $H_1$ in terms of some real fermions $\tilde \gamma_i$ that are linear combination of all $\gamma_i$, so that for $\alpha=0$ the connectivity of the fermions $\tilde \gamma_i$ has maximum exponent, achieving the case $\alpha=q$. 
For instance, the Hamiltonian $H_1 = i N \sum_{i,j} x_{ij}J_{ij} \gamma_i \gamma_j$ with $J_{ij}\sim \mathcal O(1)$ and $\alpha=0$, i.e., $p\sim 1/N^2$, can be expressed in terms of the fermions $\tilde \gamma_k = \sum_i W_{ki} \gamma_i$ with $W_{ki}\sim \mathcal O(1/\sqrt{N})$ as
$H_1 = i  \sum_{k,l} \tilde J_{kl} \tilde \gamma_k \tilde \gamma_l$, where $\tilde J_{kl}=N \sum_{i,j} W_{ki} x_{ij}J_{ij}W_{jl} \sim \mathcal O(1)$.
Basically, this behavior in the function of $\alpha$ is trivially due to the rescaling by a factor $M$ of the energy, which gives a strongly imprint for low connectivity.
Interestingly, this proves that the bound in Eq.~\eqref{eq.bound} cannot be saturated with systems of fermions except for when the connectivity exponent $\alpha$ is maximum or minimum.



}

\section{conclusion}\label{sec.conclusion}
Recently, the charging of quantum batteries has received a large attention from the scientific community. Here, we performed some analytical calculations with the aim to understand the charging of these quantum batteries through {\color{black}fermionic interactions with disorder, showing that there exist different scalings for the quantum advantage defining different classes of interactions}. Thus, we focused on a sparse SYK model, and we analyzed the resulting quantum advantage{\color{black}, explaining its origin by using a simplified model}. We determined the exact form of the quantum advantage scaling exponent, which is $a=(\alpha-q)/2+1$, for $\alpha\geq q/2$, where $\alpha$ is related to the connectivity, i.e., how much the interactions are sparse, and $q$ to the order of the interaction, i.e., the number of points involved. From the found expression of $a$, we see that as $\alpha$ approaches to $q$, the scaling exponent reaches the maximum value of $1$, 
and we get the best possible performance for this charging process.
Interestingly, we can see how, while more connectivity enhances the charging performance, an increasing of $q$ reduces it if $\alpha$ remains constant. In contrast, we get $a=1-\alpha/2$ for $q/2>\alpha\geq 0$, thus the scaling exponent reaches the maximum value of $1$ when the connectivity {\color{black}exponent} decreases to zero, $\alpha=0$. 
In conclusion, we hope that our results can be useful to clarify the charging of quantum batteries when SYK interactions are employed, at least from a theoretical point of view, and how improved performance can be achieved with many-body interactions.
In particular, batteries made with fermions showing quenched disorder in the interactions seem to be indispensable in order to get the scaling achieved for the quantum advantage. 

%

\appendix
\section*{Erratum}

In the original paper, actually in order to get $\tau^{\sharp} \sim 1/\Delta H_1$ for $H_1 = i \sum_{i>j} J_{ij} \gamma_i \gamma_j$,
where $J_{ij}\sim \mathcal O(1)$ is random with zero average to get the optimal scaling $\Delta E_1 \sim N$,
the sufficient condition  $\langle H_1h_iH_1\rangle_0\sim \Delta H_1^2$ is satisfied for a number $N'\sim N$ of indices $i$ and not all $i$, since $\sigma^y_i H_1 = - H_1 \sigma^y_i$ is not exactly satisfied as noted in the paper.
Of course it is enough that  $\langle H_1h_iH_1\rangle_0\sim \Delta H_1^2$ is satisfied for a number $N'\sim N$ of indices $i$ to get $\tau^{\sharp} \sim 1/\Delta H_1$.
To show that the above sufficient condition is satisfied in this case we can proceed as follows. The Hamiltonian $H_1$ is something like $H_1\sim\sum_{i<j}J_{ij}h_{ij}$, where
\begin{equation}
h_{ij} = \sigma^\alpha_i \left(\prod_{i<l<j}\sigma^z_l\right) \sigma^\beta_j\,,
\end{equation}
with $\alpha,\beta=x,y$. We get $\sigma^y_i h_{lk} = - h_{lk} \sigma^y_i$ if $l<i<k$ and thus there are $N_i\sim (i-1)(N-i)$ terms $h_{lk}$ that anticommutate with $\sigma^y_i$. Instead, the other $N(N-1)/2-N_i$ terms $h_{lk}$ commutate with $\sigma^y_i$. Then, we get $\sigma^y_i H_1 = - H_{1,i}\sigma^y_i+(H-H_{1,i})\sigma^y_i$, where $H_{1,i}$ describes a block having $N_i$ terms $h_{lk}$ centered around the i-th cell, from which  $\langle H_1h_iH_1\rangle_0 =  2\epsilon_0 \langle H_1 H_{1,i}\rangle_0$ for $h_i=\epsilon_0(\sigma^y_i+1)$, by noting that $\sigma^y_i\ket{0}=-\ket{0}$. Since the whole Hamiltonian $H_1$ has $N(N-1)/2$ terms $h_{ij}$, if the number of the terms $h_{lk}$ in the block $H_{1,i}$ scales as $N_i\sim N^2$, then of course we get $\langle H_1 H_{1,i}\rangle_0\sim \langle H_1^2\rangle_0$ for this block and thus $\langle H_1h_iH_1\rangle_0 \sim \Delta H_1^2$. Furthermore, in general we have $N'$ cells such that $N_i\sim N^2$, and we get that $N'$ scales linearly with $N$, i.e., $N'\sim N$, since $\sum_{i=1}^N N_i \sim N^3$, which proves that the sufficient condition  $\langle H_1h_iH_1\rangle_0\sim \Delta H_1^2$ is satisfied for a number $N'\sim N$ of indices $i$.


\begin{thebibliography}{99}
\bibitem{campaioli23} F. Campaioli, S. Gherardini, J. Q. Quach, M. Polini, and G. M. Andolina, arXiv:2308.02277 (2023).

\bibitem{Allahverdyan04} A. E. Allahverdyan, R. Balian, and T. M. Nieuwenhuizen, Europhys. Lett. 67, 565 (2004).

\bibitem{Alicki13} R. Alicki, and M. Fannes, Phys. Rev. E 87, 042123 (2013).


\bibitem{Hovhannisyan13} K. V. Hovhannisyan, M. Perarnau-Llobet, M. Huber, and A. Acin, Phys. Rev. Lett. 111, 240401 (2013).


\bibitem{Francica17} G. Francica, J. Goold, F. Plastina, and M. Paternostro, npj Quantum Inf. 3, 12 (2017).
\bibitem{Bernards19} F. Bernards, M. Kleinmann, O. G\"{u}hne, and M. Paternostro, Entropy 21, 771 (2019).

\bibitem{Touil22} A. Touil, B. \c{C}akmak, and S. Deffner, J. Phys. A: Math. Theor. 55, 025301 (2022).

\bibitem{Francica22} G. Francica, Phys. Rev. E 105, L052101 (2022).

\bibitem{Francica20} G. Francica, F. C. Binder, G. Guarnieri, M. T. Mitchison, J. Goold, and F. Plastina, Phys. Rev. Lett. 125, 180603 (2020).




\bibitem{Francica24} G. Francica, and L. Dell'Anna, Phys. Rev. E 109, 044119 (2024).

\bibitem{Francica242} G. Francica, and L. Dell'Anna, Phys. Rev. A 109, 052221 (2024).


\bibitem{Francicaqp22} G. Francica, Phys. Rev. E 105, 014101 (2022).
\bibitem{Francicaqp222} G. Francica, Phys. Rev. E 106, 054129 (2022).
\bibitem{Francicaqp23} G. Francica, and L. Dell'Anna, Phys. Rev. E 108, 014106 (2023).


\bibitem{Gyhm24} J.-Y. Gyhm, and U. R. Fischer, AVS Quantum Sci. 6, 012001 (2024).
\bibitem{Gyhm242} J.-Y. Gyhm, D. Rosa, and D. \v{S}afr\'{a}nek, Phys. Rev. A 109, 022607 (2024).


\bibitem{Binder15} F. Binder, S. Vinjanampathy, K. Modi, and J. Goold, New J. Phys. 17, 075015 (2015).

\bibitem{Campaioli17} F. Campaioli, F. A. Pollock, F. C. Binder, L. C\'{e}leri, J. Goold, S. Vinjanampathy, and K. Modi, Phys. Rev. Lett. 118, 150601
(2017).

\bibitem{Gyhm22} J.-Y. Gyhm, D. \v{S}afr\'{a}nek, and D. Rosa, Phys. Rev. Lett. 128, 140501 (2022).



\bibitem{Julia-Farre20} S. Julia-Farre, T. Salamon, A. Riera, M. N. Bera, and M. Lewenstein, Phys. Rev. Research 2, 023113 (2020).

\bibitem{Ferraro18} D. Ferraro, M. Campisi, G. M. Andolina, V. Pellegrini, and M. Polini, Phys. Rev. Lett. 120, 117702 (2018).

\bibitem{Andolina19} G. M. Andolina, M. Keck, A. Mari, M. Campisi, V. Giovannetti, and M. Polini, Phys. Rev. Lett. 122, 047702 (2019).

\bibitem{Andolina18} G. M. Andolina, D. Farina, A. Mari, V. Pellegrini, V. Giovannetti, and M. Polini, Phys. Rev. B 98, 205423 (2018).

\bibitem{Farina19} D. Farina, G. M. Andolina, A. Mari, M. Polini, and V. Giovannetti, Phys. Rev. B 99, 035421 (2019).
\bibitem{Zhang19} Y. Y. Zhang, T. R. Yang, L. Fu, and X. Wang, Phys. Rev. E 99, 052106 (2019).
\bibitem{Andolina192} G. M. Andolina, M. Keck, A. Mari, V. Giovannetti, and M. Polini, Phys. Rev. B 99, 205437 (2019).
\bibitem{Crescente20} A. Crescente, M. Carrega, M. Sassetti, D. Ferraro, Phys. Rev. B 102, 245407 (2020).

\bibitem{Rossini19} D. Rossini, G. M. Andolina, and M. Polini, Phys. Rev. B 100, 115142 (2019).


\bibitem{Rossini20} D. Rossini, G. M. Andolina, D. Rosa, M. Carrega, and M. Polini, Phys. Rev. Lett. 125, 236402 (2020).
\bibitem{Rosa20} D. Rosa, D. Rossini, G. M. Andolina, M. Polini, and M. Carrega, J. High Energ. Phys. 2020, 67 (2020).
\bibitem{Carrega21} M. Carrega, J. Kim, and D. Rosa, Entropy 2021, 23(5), 587 (2021).

\bibitem{SachdevYe} S. Sachdev and J. Ye, Phys. Rev. Lett. 70, 3339 (1993).

\bibitem{kitaevtalk} A. Kitaev, "A simple model of quantum holography." http://online.kitp.ucsb.edu/online/entangled15/kitaev/, http://online.kitp.ucsb.edu/online/entangled15/kitaev2/. Talks at KITP, April 7, 2015 and May 27, 2015.

\bibitem{Chowdhury22} D. Chowdhury, A. Georges, O. Parcollet, and S. Sachdev, Rev. Mod. Phys. 94, 035004 (2022).

\bibitem{Patel19} A. A. Patel, S. Sachdev, Phys. Rev. Lett. 123, 066601 (2019).

\bibitem{Sachdev23} S. Sachdev, arXiv:2305.01001 (2023).

\bibitem{kitaev18} A. Kitaev, S. J. Suh, JHEP 05, 183 (2018).

\bibitem{maldacena16} J. Maldacena, D. Stanford, Phys. Rev. D 94, 106002 (2016).

\bibitem{sarosi19} G. S\'{a}rosi, arXiv:1711.08482 (2019).

\bibitem{Franz18} M. Franz, M. Rozali, Nature Reviews Materials 3, 491-501 (2018).

\bibitem{hauke}  P. Uhrich, S. Bandyopadhyay, N. Sauerwein, J. Sonner, J-P. Brantut, P. Hauke, arXiv:2303.11343.

\bibitem{song17} X.-Y. Song, C.-M. Jian, and L. Balents, Phys. Rev. Lett. 119, 216601 (2017).
\bibitem{davison} R. A. Davison, W. Fu, A. Georges, Y. Gu, K. Jensen, and S. Sachdev, Phys. Rev. B 95, 155131 (2017)
\bibitem{Gnezdilov18} N. V. Gnezdilov, J. A. Hutasoit, and C. W. J. Beenakker, Phys. Rev. B 98, 081413(R) (2018).
\bibitem{Can19} O. Can, E. M. Nica, and M. Franz, Phys. Rev. B 99, 045419 (2019).
\bibitem{Guo20} H. Guo, Y. Gu, S. Sachdev, Annals of Physics 418, 168202 (2020).
\bibitem{Kulkarni22} A. Kulkarni, T. Numasawa, and S. Ryu, Phys. Rev. B 106, 075138 (2022).

\bibitem{Hosseinabadi23} H. Hosseinabadi, S. P. Kelly, J. Schmalian, and J. Marino, arxiv:2306.03898 (2023).

\bibitem{Almheiri19} A. Almheiri, A. Milekhin, and B. Swingle, arXiv:1912.04912.
\bibitem{Zhang19} P. Zhang, Phys. Rev. B 100, 245104 (2019).

\bibitem{Cheipesh21} Y. Cheipesh, A. I. Pavlov, V. Ohanesjan, K. Schalm, and N. V. Gnezdilov, Phys. Rev. B 104, 115134 (2021).

\bibitem{Francica23} G. Francica, M. Uguccioni, and L. Dell'Anna, Phys. Rev. B 108, 165106 (2023).


\bibitem{Eberlein17} A. Eberlein, V. Kasper, S. Sachdev, and J. Steinberg, Phys. Rev. B 96, 205123 (2017).


\bibitem{Zhou20} T.-G. Zhou, and P. Zhang, Phys. Rev. B 102, 224305 (2020).
\bibitem{Kruchkov20} A. Kruchkov, A. A. Patel, P. Kim, and S. Sachdev, Phys. Rev. B 101, 205148 (2020).
\bibitem{Altland19} A. Altland, D. Bagrets, and A. Kamenev, Phys. Rev. Lett. 123, 226801 (2019).


\bibitem{swingletalk} B. Swingle, ``Sparse Sachdev-Ye-Kitaev model,'' Talk at Simons Center Conference on Application of Random Matrix Theory to Many-Body Physics, Stony Brook, September 15 - September 20, 2019.
\bibitem{Garcia21} A. M. Garc\'{i}a-Garc\'{i}a, Y. Jia, D. Rosa, and J. J. M. Verbaarschot,  Phys. Rev. D 103, 106002 (2021).
\bibitem{Caceres21}  E. C\'{a}ceres, A. Misobuchi, and A. Raz,  J. High Energ. Phys. 11, 015 (2021).
\bibitem{Caceres22} E. C\'{a}ceres, A. Misobuchi, and A. Raz,  J. High Energ. Phys. 2022, 236 (2022).
\bibitem{Tezuka23} M. Tezuka, O. Oktay, E. Rinaldi, M. Hanada, and F. Nori, Phys. Rev. B 107, L081103 (2023).
\bibitem{Herasymenko23} Y. Herasymenko, M. Stroeks, J. Helsen, and B. Terhal,  Quantum 7, 1081 (2023).



\bibitem{Anandan90} J. Anandan, and Y. Aharonov, Phys. Rev. Lett. 65, 1697 (1990).


\bibitem{Bathia00} R. Bhatia, and C. Davis, The American Mathematical Monthly 107 (4): 353-357 (2000).

\bibitem{Lieb61} E. Lieb, T. Schultz, and D. Mattis,  Ann. Phys. (NY) 16, 407 (1961).

\end{thebibliography}
\end{document}